\documentclass[prd,11pt,tightenlines,nofootinbib,superscriptaddress]{revtex4}
\usepackage{amsfonts,amssymb,amsthm,bbm,amsmath}
\usepackage{verbatim}
\usepackage{graphicx}
\usepackage{color}
\usepackage{tikz}
\usetikzlibrary{calc,decorations.markings}

\newcommand{\C}{{\mathbb C}}
\newcommand{\N}{{\mathbb N}}
\newcommand{\R}{{\mathbb R}}
\newcommand{\Z}{{\mathbb Z}}

\newcommand{\one}{\mathbbm{1}}

\def\CP{{\C\mathbb{P}}}

\newcommand{\be}{\begin{equation}}
\newcommand{\ee}{\end{equation}}
\newcommand{\beq}{\begin{eqnarray}}
\newcommand{\eeq}{\end{eqnarray}}
\newcommand{\bea}{\begin{eqnarray}}
\newcommand{\eea}{\end{eqnarray}}
\newcommand{\nn}{\nonumber}

\newcommand{\bra}{\langle}
\newcommand{\ket}{\rangle}

\newcommand{\rd}{\mathrm{d}}

\newcommand{\bpm}{\begin{pmatrix}}
\newcommand{\epm}{\end{pmatrix}}
\newcommand{\bvm}{\begin{vmatrix}}
\newcommand{\evm}{\end{vmatrix}}

\def\nn{\nonumber}

\begin{document}

\title{Spin Foams Without Spins}

\author{{\bf Jeff Hnybida}}
\affiliation{Institute  for  Mathematics,  Astrophysics,  and  Particle  Physics  Faculty  of  Science,
Radboud  University,  Nijmegen,  Netherlands}
\email{j.hnybida@science.ru.nl}

\begin{abstract}
We formulate the spin foam representation of discrete SU(2) gauge theory as a product of vertex amplitudes each of which is the spin network generating function of the boundary graph dual to the vertex.  In doing so the sums over spins have been carried out.  The boundary data of each $n$-valent node is explicitly reduced with respect to the local gauge invariance and has a manifest geometrical interpretation as a framed polyhedron of fixed total area.  Ultimately, sums over spins are traded for contour integrals over simple poles and recoupling theory is avoided using generating functions.
\end{abstract}

\maketitle


\section{Introduction}

Lattice gauge theory is an ultraviolet regularization of Quantum Gauge Theory \cite{Drouffe:1983fv}.  It has a finite number of degrees of freedom which are the holonomies around ordered plaquettes of the lattice taken to be elements of the gauge group, usually a special unitary group.  

The dual of a Lattice Gauge theory is known as a spin foam model \cite{Oeckl:2000hs} in which the integrals over group variables are traded for sums over unitary irreducible representations of the gauge group.  This duality is the result of the well-known character expansion technique of Boltzmann weights \cite{Drouffe:1983fv}.  Note that spin foam models are usually used to study state sum models of quantum gravity, which involve extra constraints, and are thus not of the Yang-Mills variety.  However, Yang-Mills is a regularization of BF theory which is usually the starting point of most spin foam models \cite{Baez:1997zt}.

The representation of Lattice Gauge Theory as a spin foam model has been revisited throughout the last few decades \cite{Aroca:1997hd,Conrady:2005qu,Cherrington:2007ax} however the analytical advantage of this representation is mixed.  While the group integrations are performed exactly, the sums over spins and the subsequent recoupling theory involved are daunting.

We take a first step in easing these difficulties by trading the sums over spins for contour integrals and avoiding recoupling theory by using generating functions.  Note that these contour integrals are localized at the vertices, as opposed to spins which are associated to the faces of the 2-complex (or plaquettes of a lattice).  We consider pure SU(2) lattice gauge theory, but extensions to the other unitary groups is expected.

To achieve this we isolate the group variables by inserting a coherent resolution of identity on $n$-valent SU(2) tensors.  The coherent representation is labelled by a set of $n$ spinors, i.e. $\C^{2n}$.  This idea is similar to other proposed methods of integrating out the group variables in lattice gauge theories \cite{Holtkamp:1981su}  and more recently \cite{Vairinhos:2014uxa}.   However, we reduce our auxiliary variables, a set of $n$ spinors, with respect to the SU(2) gauge invariance to the Grassmannian subspace $\text{Gr}(2,n) \cong \C^{2n}//\text{SU}(2)$.  These states were first studied by Fujii et. al. \cite{Fujii:1995ds} and later in the context of spin foam models by Freidel and Livine \cite{Freidel:2010tt}.  These elements of the Grassmannian are scale invariant, have a canonical action of the conformal group, and a nice geometrical interpretation as framed polyhedra \cite{Bianchi:2010gc,Livine:2013tsa}.

Moreover, and maybe most importantly, these auxiliary variables factorize at the vertices as the generating functions of spin network graphs dual to the vertices.  These generating functions have a simple form in terms of Pl\"ucker coordinates of the Grassmannian elements arranged in cycles of the graph \cite{westbury1998generating,Freidel:2012ji,Bonzom:2012bn}.

The plan of the paper is as follows:  First we review SU(2) lattice gauge theory with the Wilson action.  Then we review the character expansion in terms of spinors in the Bargmann-Fock representation which explains the origin of the U($n$) projector.  We then review some facts about the U($n$) projector, in particular the covariance with respect to GL(2,$\C$).  We then present a new representation of the projector by the introduction of a contour integral.  Using this contour integral representation we discuss the contraction of a set of projectors into a vertex amplitude as well as the defining identities of BF and Yang-Mills gauge theories: the loop identity.  Finally, we perform the complex quotient $\C^{2n}/\text{GL}(2,\C) \cong \text{Gr}(2,n)$ of the projector to the Grassmannian subspace. 
 

\section{SU(2) Lattice Gauge theory}

The simplest action for pure lattice gauge theory is the Wilson action and is defined in terms of a set of group elements $g_e \in \text{SU(2)}$ living on the edges of a 2-complex:\footnote{A 2-complex is a generalization of a graph which is a 1-complex.  It is simply a collection of vertices, edges, and faces.}
\be
  S(g_e) = \sum_{f} \text{tr}(G_f) \qquad G_f = g_{e_{1}} g_{e_{2}} \cdots g_{e_{k}} 
\ee
where the sum is over faces (2-cells) of the 2-complex.  Here $G_f$ is the product of group elements around the face $f$.  For simplicity one can assume that the vertices and edges of this 2-complex are defined by a lattice and the faces by the plaquettes, however the results presented here are applicable for general complexes.

The Lattice Gauge theory partition function is then
\be \label{eqn_YM_part_func}
  Z_{\text{YM}}(\beta) = \int \rd g_e  \, e^{\frac{\beta}{2} S(g_e)}
\ee 
where $\rd g_e$ is the Haar measure and $\beta$ is related to the Yang-Mills coupling constant and the lattice spacing.  The relationship with the continuum path integral is found by taking a naive continuum limit \cite{Drouffe:1983fv}.  The limit $\beta \rightarrow \infty$ of (\ref{eqn_YM_part_func}) is a well known topological field theory known as BF theory:
\be \label{eqn_BF_limit}
  \lim\limits_{\beta \rightarrow \infty} Z_{\text{YM}}(\beta) \propto \int \rd g_e  \, \prod_{f} \delta(G_f)
\ee

Hence (\ref{eqn_YM_part_func}) is a regularization of the BF theory partition function (\ref{eqn_BF_limit}) which is an ill-defined product of distributions.  

While it is probably impossible to evaluate (\ref{eqn_YM_part_func}) exactly for non-trivial lattices, it is possible to perform the group integrals by introducing coherent states, i.e. spinors, living at the vertices.  This is the spirit of the methods employed in spin foam models of quantum gravity \cite{Baez:1997zt}, \cite{Livine:2007vk}.  

To see how this is done, first recall the character expansion \cite{Drouffe:1983fv} for a plaquette of the Wilson action
\be \label{eqn_wilson_char_exp} 
 \frac{\beta}{2} e^{\frac{\beta}{2} \text{tr}(G)} = \sum_{j \in \N/2} (2j+1)I_{2j+1}(\beta) \chi_j(G)
\ee 
which can be derived using the generating series of the modified Bessel function and the Weyl character formula.  The modified Bessel function acts as an effective heat kernel because for large $j$
\be \label{eqn_heat_kernel_approx}
  \frac{I_{2j+1}(\beta)}{I_{1}(\beta)} \approx e^{-\frac{2j(j+1)}{\beta}}
\ee
and so the character expansion of the delta function 
\be \label{eqn_delta_char}
  \delta(G) = \sum_{j \in \N/2} (2j+1) \chi_{j}(G)
\ee
is recovered in the limit $\beta \rightarrow \infty$.  The product of group elements in each character can be factorized by resolutions of identity in the standard orthonormal basis and the group integrals can be performed.  For the triangulation of a 3-manifold without boundary the result is 
\be \label{eqn_YM_6j}
  Z_{\text{YM}}(\beta) = \sum_{j_f \in \N/2} (-1)^\chi \prod_{f} (2j_f+1)I_{2j_f+1}(\beta) \prod_{v} \{6j\}
\ee
where $\{6j\}$ is the normalized Wigner 6j symbol and the sign can be shown to be $\chi = \sum_f j_f + \sum_e \sum_{f \cap e} j_f$ \cite{Barrett:2008wh}. 

The expansion (\ref{eqn_YM_6j}) is a regularization of the Ponzano-Regge model which is the prototypical spin foam model, but is generically divergent for $\beta \rightarrow \infty$.  The insertion of various weights into the SU(2) delta corresponding to observables in the Ponzano-Regge model were also studied in \cite{Livine:2008sw}.

Instead of using orthonormal states, in Section \ref{sec_coh_states} we use SU(2) coherent states to expand and factorize all of the $g_e$ in the characters.  For more details see \cite{Banburski:2014cwa}.  We do this in order to take advantage of the geometrical properties of the coherent states, and to use generating functions for vertex amplitudes, rather than recoupling coefficients such as the $\{6j\}$ symbol.

\section{Bargmann-Fock Space} \label{sec_coh_states}

In this section we show how to expand the SU(2) characters as an integral over spinors.  The main identity we require to do this is the reproducing kernel on spinor space
\be \label{eqn_rep_kern_id}
  \int_{\C^2} \rd \mu(x) \, \frac{\bra z|x\ket^{k} \bra x | w\ket^{k'}}{k!k'!} = \frac{\bra z|w\ket^{k}}{k!} \delta_{kk'} \quad\qquad \rd \mu(x) = \frac{\rd^4 x}{\pi^2} e^{-\bra x | x \ket} 
\ee
where we represent covariant/contravariant spinors $z,w \in \C^2$ by row/column vectors
\begin{align}
&\bra z | = \bpm \bar{z}^0 & \bar{z}^{1} \epm \qquad |w\ket = \bpm w^0 \\ w^1 \epm \nonumber \qquad [ w | = \bpm -w^1 & w^{0} \epm \qquad |z] = \bpm -\bar{z}^1 \\ \bar{z}^0 \epm \nonumber \\  
&\bra z | w\ket = \bpm \bar{z}^0 & \bar{z}^{1} \epm\bpm w^0 \\ w^1 \epm = \delta_{AB} \bar{z}^{A} w^{B}  \qquad [z|w\ket = \bpm -z^1 & z^{0} \epm\bpm w^0 \\ w^1 \epm = \epsilon_{AB} z^{A} w^{B} \nonumber 
\end{align}
and we denote $|\check{z}\ket = |z]$ which is an anti-involusion $\check{\check{z}} = -z$.  The identity (\ref{eqn_rep_kern_id}) is derived by expanding both sides of the following Gaussian integral
\be
  \int_{\C^2} \rd \mu(x) \, e^{\bra z|x\ket +  \bra x | w\ket} = e^{ \bra z | w \ket}
\ee
This is in fact enough to show that $e^{\bra z | w \ket}$ is the reproducing kernel on the space of holomorphic functions on $\C^2$
\be \label{eqn_rep_kern}
  \int_{\C^2} \rd \mu(z) e^{\bra z|w\ket} f(z) = f(w) \qquad \forall f \in L^2_{\text{hol}}(\C^2,\rd \mu)
\ee  

The space $L^{2}_{\text{hol}}(\C^2,\rd\mu)$ is a representation space for SU(2) known as the Bargmann-Fock space \cite{Bargmann, Schwinger}.  The relation (\ref{eqn_rep_kern_id}) demonstrates the coherent resolution of the identity on the representation of spin $j=k/2$.  Hence we have the coherent expansion of the spin-$j$ character 
\be \label{eqn_coherent_character}
  \chi_j(G) = \int_{\C^2} \rd \mu(z) \frac{\bra z | g_{e_1} g_{e_2} \cdots g_{e_{k}} | z \ket^{2j}}{(2j)!}
\ee
Using (\ref{eqn_rep_kern_id}) we can factorize all of the $g_e$ by inserting resolutions of the identity.  If there are $n$ faces containing the edge $e$ then there will be $n$ pairs of spinors $z^{e}_f, w^{e}_f$ contracting $g_e$ for each face $f$, with the condition that if two edges $e$, $e'$ are adjacent and share the face $f$ then $w^{e}_f = z^{e'}_f$.  Hence we have
\be \label{eqn_YM_expanded}
  Z_{\text{YM}}(\beta) = \int_{\C^{2EF}} \rd \mu(z^{e}_{f}) \prod_f \sum_{j_f \in \N/2} (2j_f+1)I_{2j_f+1}(\beta)  \prod_e \int_{\text{SU(2)}} \rd g_e \, \prod_{f\in e}\frac{\bra z^{e}_{f} | g_e | w^{e}_{f} \ket^{2j_f}}{(2j_f)!}
\ee
The focus of this paper is on these group averaged reproducing kernels. Due to the orthogonality (\ref{eqn_rep_kern_id}) we can study the exponentiation of these kernels, i.e. sum over spins without loss of generality, which we do in the next section.

\section{The U(n) Coherent Projector}

The group averaged reproducing kernels in (\ref{eqn_YM_expanded}) project holomorphic functions onto the SU(2) invariant subspace of the Bargmann-Fock space of $n$ spinors $L^{2}_{\text{hol}}(\C^{2n},\rd\mu(z_i))$ where $\rd\mu(z_i) \equiv \rd\mu(z_1)\rd\mu(z_2)\cdots \rd\mu(z_n)$.  Indeed, given $2n$ spinors $\{z_i\}$, $\{w_i\}$ define
\be \label{eqn_proj_group}
  P(z_i,w_i) \equiv \int_{\text{SU(2)}} \rd g \, e^{\sum_i \bra z_i | g_ | w_i \ket}
\ee
Then it is easy to show using (\ref{eqn_rep_kern}) that for $f \in L^{2}_{\text{hol}}(\C^{2n},\rd\mu(z_i))$
\be
  \int_{\C^{2n}} \rd \mu(z_i) P(z_i,w_i) f(z_i) = \int_{\text{SU(2)}} \rd g \,f(g w_i)
\ee
which is invariant under rigid SU(2) transformations.  Also $P\circ P = P$ where convolution is with respect to the measure $\rd\mu(z_i)$.

The space $L^{2}_{\text{hol}}(\C^{2n},\rd\mu(z_i))$ has a natural action of $U \in \text{U(n)}$ given by
\be
  \hat{U} f(z_i) = f( (U z)_i) \qquad (Uz)_i = \sum_j U_{ij} z_j
\ee
and holomorphicity extends this action to SL(n,$\C$).  The identity on $L^{2}_{\text{hol}}(\C^{2n},\rd\mu(z_i))$ is simply the reproducing kernel
\be \label{eqn_rep_kern_triv}
  \one_{n}(z_i,w_i) \equiv e^{\sum_{i=1}^{n} \bra z_i | w_i \ket}
\ee
while (\ref{eqn_proj_group}) is the group average of $\one_{n}(z_i,w_i)$  and is thus the projector on the symplectic subspace $\C^{2n}//SU(2) \cong \text{Gr}(2,n)$.

As shown in \cite{Freidel:2010tt} the group integration in (\ref{eqn_proj_group}) can be performed resulting in 
\be \label{eqn_proj_first}
  P(z_i,w_i) = \sum_{J=0}^{\infty} \frac{(z_i|w_i)^J}{J!(J+1)!} 
\ee
where we denote the Hermitian scalar product $(z_i|w_i) \equiv \sum_{1\leq i<j \leq n} \bra z_j|z_i][w_i|w_j\ket$.  

The states for fixed $J$ were first derived by Fujii et. al. in \cite{Fujii:1995ds} as U(n) coherent states of highest weight $[J,J,0,...,0]$ on the Grassmannian Gr(2,n).  Later these coherent states were rediscovered in the context of Loop Quantum Gravity by Freidel and Livine \cite{Freidel:2010tt} where $J$ was given the interpretation of the total area of a closed polyhedron with $n$ faces embedded into $\R^3$.

This scalar product $(z_i|w_i)$ is U(n) invariant and is homogeneous with respect to the diagonal GL(2,$\C$) action
\be
  (gz_i | gw_i) = |\det(g)|^2 (z_i|w_i) \qquad \forall g \in \text{GL}(2,\C)
\ee
As shown in both \cite{Fujii:1995ds} and \cite{Freidel:2010tt} the symplectic reduction to the Grassmannian can be achieved by a simple complex quotient $\text{Gr}(2,n) \cong \C^{2n}/\text{GL}(2,\C)$.  We will also follow the same method in Section \ref{sec_closure} using the form of the projector which we introduce in the next section.

To close this section we remark that the partition function in (\ref{eqn_YM_expanded}) can be expressed in terms of the projectors (\ref{eqn_proj_group}) as
\be \label{eqn_YM_projectors}
  Z_{\text{YM}}(\beta) = \int_{\C^{2EF}} \rd \mu(z^{e}_{f}) \prod_f \delta_{YM}(z^{\tilde{e}}_{f},w^{\tilde{e}}_f,\beta)  \prod_e P(z^{e}_f,w^{e}_f)
\ee
where $\delta_{YM}$ is an integral kernel, which we will define in section \ref{sec_YM_loop} and $\tilde{e}$ is any edge contained in $f$.  There we will show that the integrals over these kernels produce the face factors $(2j_f+1)I_{2j_f+1}(\beta)$ in (\ref{eqn_YM_expanded}).  Again note that $w^{e}_f = z^{e'}_f$ if $e$, $e'$ are adjacent and share the face $f$.

\section{An Integral Representation of the Projector}

Let us begin with the projector in the form (\ref{eqn_proj_first}).  An integral representation of the reciprocal Gamma function, due to Hankel, is given by
\be
  \frac{1}{\Gamma(z)} = \frac{1}{2\pi i} \oint_{\gamma_0} \rd s \, s^{-z-1} e^{s}
\ee
where $\gamma_0$ is a contour starting at infinity below the negative real axis, encircling the origin and ending at infinity above the real axis, known as a Hankel (keyhole) or Bromwich contour depending on the branch points.  See Figure \ref{fig_contour}.

\tikzset{arrow data/.style 2 args={%
      decoration={%
         markings,
         mark=at position #1 with \arrow{#2}},
         postaction=decorate}
      }%

\begin{figure} 
\begin{tikzpicture}
\draw [help lines,->] (-3.25, 0) -- (3.25,0);
\draw [help lines,->] (0, -3.25) -- (0, 3.25);
\draw[very thick,arrow data={0.25}{stealth},arrow data={0.75}{stealth},scale=1.5,domain=-3.141/2+0.45:3.141/2-0.45,smooth,variable=\t] plot ({-1/cos(\t r)+2},{sin(\t r)/cos(\t r)});

\node at (3.6,-0.4){$\text{Re}(s)$};
\node at (0.64,3.53) {$\text{Im}(s)$};
\node at (-0.4,-0.43) {$O$};
\node at (1.6,1.6) {$\gamma_{0}$};
\end{tikzpicture}
\caption{The contour $\gamma_0$ encircling the origin in a counter-clockwise manner.  If the integrand is analytic in $s$ then $\gamma_0$ can be closed to the unit circle, or depending on the branch points one may use a Hankel (keyhole) or Bromwich (inverse Laplace transform) contour. } \label{fig_contour}
\end{figure}
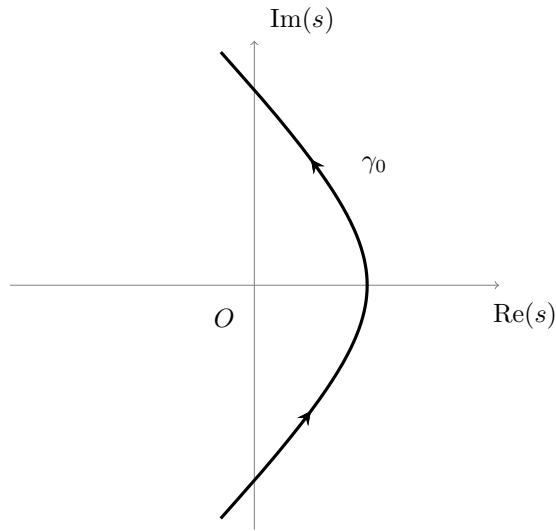

Whenever the integrand is analytic in $s$ (except at the origin) we can deform $\gamma_0$ to any closed contour encircling the origin in the counterclockwise direction.  Now we can represent the $1/(J+1)!$ factor in the projector by this integral expression
\be \label{eqn_proj}
  P(z_i,w_i) = \frac{1}{2\pi i}\oint_{\gamma_0} \frac{\rd s}{s^2} \, e^{s + \frac{1}{s}(z_i|w_i)}
\ee
The projector is in fact a hypergeometric series denoted $P(z_i,w_i) = {}_{0}F_{1}(;2;(z_i|w_i))$ which is of the Bessel variety and (\ref{eqn_proj}) is one of its integral representations.  One may also view this function as an inverse Laplace transform $P(z_i,w_i) = \mathcal{L}^{-1}\{ e^{\frac{1}{s}(z_i|w_i)}/s^2 \}(1)$ where the dual Laplace variable is evaluated at unity.

One can check that the projector equals its square as follows
\begin{align} \label{eqn_proj_squared}
  \int_{\C^{2n}} \rd\mu(x_i) P(z_i,x_i) P(x_i,w_i) 
  &= \frac{1}{(2\pi i)^2}\oint_{\gamma_0} \frac{\rd s}{s^2} \oint_{\gamma_0} \frac{\rd t}{t^2} \, e^{s + t} \int \rd\mu(x_i) e^{\frac{1}{s}(z_i|x_i) +  \frac{1}{t}(x_i|w_i)}  \\
  &= \frac{1}{(2\pi i)^2} \oint_{\gamma_0} \oint_{\gamma_0} \rd s \, \rd t \, 
  \frac{e^{s + t}}{\left(st - (z_i|w_i)\right)^2} 
\end{align}
where in the second line we used the formula (see \cite{Freidel:2012ji} for proof)
\be
  \int_{\C^{2n}} \rd\mu(x_i) e^{ a(z_i|x_i) + b(x_i|w_i)}
  = \left(1 - ab(z_i|w_i)\right)^{-2}
\ee
Now making a change of variable $t \rightarrow t + (z_i|w_i)/s$ gives the desired result
\be 
  \int_{\C^{2n}} \rd\mu(x_i) P(z_i,x_i) P(x_i,w_i) 
  = \frac{1}{(2\pi i)^2} \oint_{\gamma_0} \rd t \, \frac{e^{t}}{t^2} \oint_{\gamma_0} \frac{\rd s}{s^2} \, e^{s + \frac{1}{s}(z_i|w_i)} = P(z_i,w_i)
\ee
since the integral over $t$ evaluates to $2\pi i$.

There are several advantages to the contour integral representation (\ref{eqn_proj}) of the projector over (\ref{eqn_proj_first}).  Firstly contour integrations are generally easier to compute that sums like in (\ref{eqn_proj_first}) or group integrals as in (\ref{eqn_proj_group}).  Secondly, the exponentiation of the Hermitian scalar product $(z_i|w_i)$ allows for the possibility of Gaussian integration which we will demonstrate in the next section.  Finally, the U(1) variables $s,t$ will be useful in computing the K\"ahler reduction of the spinors to the Grassmannian.  Their physical interpretation and the implications of scale invariance will be explored elsewhere \cite{Livine_to_come}.

\section{The Vertex Amplitude}
\label{sec_vertex_amp}

The advantage of using the contour integral representation (\ref{eqn_proj}) of the projector is that the Hermitian scalar product $(z_i|w_i)$ is Gaussian and can thus be integrated easily as was shown in \cite{Freidel:2012ji}.

This Gaussian integral takes place at each vertex of the 2-complex.  In order to factorize the spinors $z^{e}_f$ at the vertices we use the property $P = P \circ P$ of the projector to express (\ref{eqn_YM_projectors}) as
\begin{align} \label{eqn_YM_vertices}
  Z_{\text{YM}}(\beta) 
  &= \int_{\C^{2EF}} \rd \mu(z^{e}_{f}) \prod_f \delta_{YM}(z^{\tilde{e}}_{f},w^{\tilde{e}}_f,\beta)  \prod_e \int_{\C^{2EF}} \rd \mu(x^{e}_{f}) P(z^{e}_f,x^{e}_f) P(x^{e}_f,w^{e}_f) \nonumber\\
  &=\int_{\C^{2EF}} \rd \mu(x^{e}_{f}) \prod_v A_{\Gamma_v}(x^{e}_f,\beta)
\end{align}
where the vertex amplitudes are defined by
\be \label{eqn_vertex_amp}
  A_{\Gamma_v}(x^{e}_f,\beta) = \int_{\C^{2EF}} \rd \mu(z^{e}_{f}) \prod_{\tilde{e}\cap v} \delta_{YM}(z^{\tilde{e}}_{f},w^{\tilde{e}}_f,\beta)  \prod_{e \cap v} P(z^{e}_f,x^{e}_f)
\ee
Note that each edge of the 2-complex is canonically oriented by the association of the holomorphic/anti-holomorphic spinors $x^{e}_{f}$.  The label $\Gamma_v$ is the graph associated to the contraction of spinors meeting at $v$.  Indeed, the spinors $z^{e}_{f}$ in (\ref{eqn_vertex_amp}) are implicitly related via the adjacency of the underlying 2-complex.  In (\ref{eqn_YM_projectors}) this was imposed formally by the condition $w^{e}_{f} = z^{e'}_{f}$ if $e', e$ are adjacent and share the same face $f$.

For simplicity, assume $\Gamma$ is a simple\footnote{The results hold for non simple graphs as well, but for simple graphs we can denote an edge by a pair of vertices which makes the notation more manageable.} directed graph with vertices labelled by $i,j,k \in \Z^V$ where $V$ is the number of vertices and edges are denoted $\bra ij\ket$.  Then as shown in \cite{Freidel:2012ji} we can perform the following Gaussian integral
\be \label{eqn_gen_func}
  \int_{\C^{2V(V-1)}} \rd \mu(w^{i}_{j}) e^{\sum_{i\in \Gamma} (w^{i}_{j}|x^{i}_{j})}\Big|_{w^{j}_{i} = \epsilon_{ij} \check{w}^{i}_{j}} = \frac{1}{\det\left( 1 + X^\Gamma \right)}
\ee
where contraction is imposed by the reality condition $w^{j}_{i} = \epsilon_{ij} \check{w}^{i}_{j}$ where $\epsilon_{ij} = \pm 1$ if $\bra ij\ket$ is positively/negatively oriented and zero if there is no edge connecting $i$ with $j$.\footnote{This reality condition can be imposed using the following integral kernel $e^{\sum_{\bra ij\ket} \epsilon_{ij} [ w^{i}_{j} | w^{j}_{i}\ket}$.}  The notation is such that the spinor $x^{i}_{j}$ is associated to vertex $i$ and directed to vertex $j$.  The Hermitian scalar product at each vertex $i$ is explicitly 
\be
  (w^{i}_{j}|x^{i}_{j}) = \sum_{\bra ij\ket \cap \bra ik\ket} \bra w^{i}_{j}|w^{i}_{k}] [x^{i}_{j}|x^{i}_{k}\ket
\ee
i.e. all pairs of edges meeting at $i$.  The matrix $X^{\Gamma}$ is of size $2V$ composed of $2\times 2$ blocks defined by the outer product of spinors
\be
  X^{\Gamma}_{ij} = \epsilon_{ij} |x^{i}_{j}\ket[x^{j}_{i}|
\ee
and has a remarkable property, referred to as the scalar loop property in  \cite{Freidel:2012ji}, which allows its determinant to be given in terms of disjoint unions of cycles on the graph $\Gamma$.  

Let us now consider the vertex amplitude (\ref{eqn_vertex_amp}) with the projectors in the form (\ref{eqn_proj}).  We neglect the kernels $\delta_{YM}$ for now as they just amount to some extra contour integrals.  By integrating over $w^{i}_{j}$ in the pattern of a graph we can use (\ref{eqn_gen_func}) to obtain a meromorphic function of the spinors $x^{i}_{j}$.  More explicitly 
\be \label{eqn_proj_cont}
  A_\Gamma(x^{i}_{j}) \equiv \int_{\C^{2V^2}} \rd \mu(w^{i}_{j}) \prod_{i\in \Gamma} P(w^{i}_{j},x^{i}_{j})\Big|_{w^{j}_{i} = \epsilon_{ij} \check{w}^{i}_{j}} = \oint_{\gamma^{V}_0} \frac{\rd s_i}{(2\pi i)^V} \frac{e^{\sum_i s_i}}{\det\left( D(s_i) + X^\Gamma \right)}
\ee
where $D(s_i) = \text{diag}(s_1,s_2,...,s_V)$.  We can similarly compute the determinant of this graph in terms of cycles. 

For example the tetrahedral vertex amplitude is given by contracting four projectors, labelled 1,...,4 in the pattern of a 3-simplex.  We choose the orientation of $\bra ij \ket$ to be positive if $i<j$.  Then the contraction of four projectors (\ref{eqn_proj_cont}) in this way has the form
\be \label{eqn_tet_gen_func}
  A_{3S}(x^{i}_{j}) = \oint_{\gamma^{4}_0} \frac{\rd s_i}{(2\pi i)^4} \frac{e^{s_1 + s_2 + s_3 + s_4}}{(s_1 s_2 s_3 s_4 - s_1 A_{234} - s_2 A_{134} -s_3 A_{124} -s_4 A_{123} +A_{1234} -A_{1342} - A_{1423})^2}
\ee
where for each cycle of the graph
\be
  A_{1 2 \dots p} = [x^{1}_{p}|x^{1}_{2}\ket[x^{2}_{1}|x^{2}_{3}\ket \cdots [x^{p}_{p-1}|x^{p}_{1}\ket
\ee
and the sign of a cycle in (\ref{eqn_tet_gen_func}) is determined by the orientation of the graph as the number of edges in the cycle that agrees with the orientation of $\Gamma$ plus one (see \cite{Freidel:2012ji}). 

The expression (\ref{eqn_tet_gen_func}) is a generating function for the 6j symbol, as shown in \cite{Bargmann}.  Indeed, expanding in a power series one finds six unrestricted sums over positive integers $2j_{ij}$ corresponding the homogeneity of the spinors $x^{i}_{j}$ and a seventh sum which defines the Racah formula of the Wigner 6j symbol.  The contour integrals ensure the proper normalization giving
\be \label{eqn_normalized_6j}
  A_{3S}(x^{i}_{j}) = \sum_{\{j_{ij}\}} (-1)^s \{6j\} \prod_i C(j_{ij},x^{i}_{j}) 
\ee
where $\{6j\}$ is the normalized 6j symbol and $C(j_{ij},x^{i}_{j})$ are normalized 3j symbols in the coherent representation.  The sign $s$ comes from the particular orientation of the graph, i.e. the relative signs in the denominator of (\ref{eqn_tet_gen_func}), which is inherited from the orientation of the 2-complex.

Using the expansion (\ref{eqn_normalized_6j}) and the orthonormality of the $C(j_{ij},x^{i}_{j})$ one could now construct the regularized Ponzano-Regge model (\ref{eqn_YM_6j}) if given a mechanism of inserting the face factors, i.e. $2j+1$ in (\ref{eqn_wilson_char_exp}) for BF theory and $(2j+1)I_{2j+1}(\beta)$ in (\ref{eqn_delta_char}) for YM theory.  We show how this can be done using integral kernels in the following sections.

Finally, we note that restricting the spinors $x^{i}_{j}$ to the Grassmannian, as we will do in Section \ref{sec_closure} does not change the form of these vertex amplitudes.  This is due to the homogeneity of the determinant in (\ref{eqn_proj_cont}) with respect to GL(2,$\C$) transformations.

\section{The BF Loop Identity} \label{sec_loop_id}

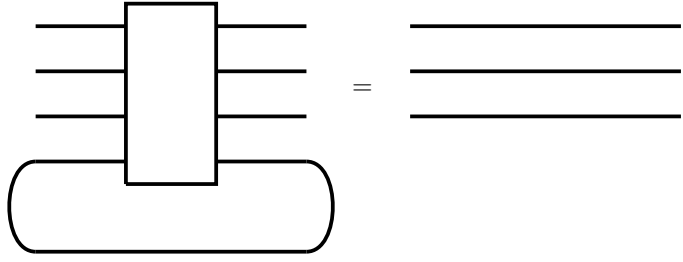
\begin{figure} 
\begin{tikzpicture}[baseline=0,scale=0.6,line width=0.50mm]
  \draw (-3,1.5) -- (-1,1.5);
  \draw (-3,0.5) -- (-1,0.5);
  \draw (-3,-0.5) -- (-1,-0.5);
  \draw (-3,-1.5) -- (-1,-1.5);
  \draw (-1,-2) --(-1,2) -- (1,2) -- (1,-2) -- (-1,-2);
  \draw (1,1.5) -- (3,1.5);  
  \draw (1,0.5) -- (3,0.5);  
  \draw (1,-0.5) -- (3,-0.5);  
  \draw (1,-1.5) -- (3,-1.5); 
  \draw (-3,-1.5) to [out=180,in=180] (-3,-3.5);
  \draw (3,-1.5) to [out=0,in=0] (3,-3.5);
  \draw (-3,-3.5) -- (3,-3.5);
\end{tikzpicture}
= \:\:\:
  \begin{tikzpicture}[baseline=0,scale=0.6,line width=0.50mm]
  \draw (-3,1.5) -- (3,1.5);
  \draw (-3,0.5) -- (3,0.5);
  \draw (-3,-0.5) -- (3,-0.5); 
\end{tikzpicture}
\caption{The graphical representation of the loop identity for a projector with four strands.  The box on the LHS represents the group averaging in (\ref{eqn_proj_group}) of the trivial reproducing kernel $\one_4$ defined in (\ref{eqn_rep_kern_triv}).  The RHS is simply $\one_3$ since the closed loop evaluates to the delta function on SU(2).   } \label{fig_loop_id}
\end{figure}

In this section we discuss the defining identity of BF theory known as the Loop identity \cite{Oeckl:2005rh}.  The loop identity, illustrated in Fig. \ref{fig_loop_id}, results from the fact that each loop of the BF theory partition function (\ref{eqn_BF_limit}) is a group delta function.  Therefore the integration over the group variable (the box in the diagram) is performed by the delta (the loop) and results in the group variable being evaluated at the identity (no box on the RHS).  

However, the trace of one strand of the projector, using the reproducing kernel $e^{\bra w_n | z_n \ket}$, does not yield the character expansion (\ref{eqn_delta_char}) since
\be \label{eqn_proj_trace}
  \int_{\C^{2}} \rd\mu(z_n) \int_{\C^{2}} \rd\mu(w_n) e^{\bra w_n | z_n \ket} P(z_i,w_i) = \int_{\text{SU}(2)} \rd g \, e^{\sum_{i=1}^{n-1} \bra z_i | g | w_i \ket} \sum_{j_n \in \N/2} \chi_{j_n}(g)
\ee
where we used the coherent expansion of the character (\ref{eqn_coherent_character}).  

In order to recover the delta function on SU(2) we would need to insert a factor of $2j_n+1$ into the sum over $j_n$.  As shown in \cite{Dupuis:2011fz} this can be achieved by the following integral identity
\be
   \int \rd \mu(z)( \bra z | z \ket -1) e^{\bra z|g|z\ket} = \sum_{j \in \N/2} (2j+1) \chi_{j}(g) = \delta(g)
\ee
Thus by replacing the reproducing kernel $e^{\bra w_n | z_n \ket}$ on the LHS of (\ref{eqn_proj_trace}) with 
\be
  \delta_{\text{BF}}(z_n,w_n) \equiv ( \bra z_n | w_n \ket -1) e^{\bra w_n | z_n\ket}
\ee
we have the following loop identity
\be \label{eqn_loop_id_beginning}
  \int_{\C^{2}} \rd \mu(z_n)\int_{\C^{2}} \rd \mu(w_n) \delta_{\text{BF}}(z_n,w_n) P(z_i,w_i) = \int_{\text{SU}(2)} \rd g \, e^{\sum_{i=1}^{n-1} \bra z_i |g| w_i \ket}\delta(g) = e^{\sum_{i=1}^{n-1} \bra z_i | w_i \ket}
\ee
which is responsible for triangulation invariance in BF theory.  We would now like to see how this identity manifests itself for the projector (\ref{eqn_proj}).

To handle the extra measure factor $( \bra z_n | z_n \ket -1)$ let us integrate with respect to the measure
\be
  \rd \mu_\sigma(z_n) = e^{ -\sigma\bra z_n | z_n \ket} \frac{\rd z_{n}^{4}}{\pi^2}
\ee
and then use the fact that
\be
  -\left( \frac{\partial}{\partial \sigma} + 1 \right) \int \rd \mu_\sigma(z_n) P(z_i,w_i) \Big|_{\sigma = 1} = \int \rd \mu(z_n)( \bra z_n | z_n \ket -1) P(z_i,w_i)
\ee
Let us use a subscript $n$ on $(z_i|w_i)_{n}$ to denote the Hermitian scalar product for $i = 1,...,n$.  Similarly let $\bra z_i,w_i \ket_n = \sum_{i=1}^{n} \bra z_i | w_i \ket$ denote the canonical Hermitian inner product on $\C^{2n}$. Then 
\begin{align}
  \int \rd \mu_\sigma(z_n) P(z_i,w_i) 
  &= \frac{1}{2\pi i} \oint_{\gamma_0} \frac{\rd s}{s^2} \, e^{s + \frac{1}{s}(z_i|w_i)_{n-1}}
  \int \frac{\rd z_{n}^{4}}{\pi^2} e^{ -\sigma\bra z_n | z_n \ket - (1/s) \sum_{j \neq n} \bra z_n | z_j \ket [w_j |z_n\ket} \nonumber \\
  &= \frac{1}{2\pi i} \oint_{\gamma_0} \frac{\rd s}{s^2} \, \frac{e^{s + \frac{1}{s}(z_i|w_i)_{n-1}}}{\det\left(\sigma + (1/s)\sum_{j \neq n} |z_j \ket [w_j| \right)} \nonumber \\
  &= \frac{1}{2\pi i} \oint_{\gamma_0} \rd s \, \frac{e^{s + \frac{1}{s}(z_i|w_i)_{n-1}}}{\sigma^2s^2-\sigma s\bra z_i,w_i \ket_{n-1}+(z_i|w_i)_{n-1}} \label{eqn_loop_poles}
\end{align}
where in the last step we used the lemma (for proof see \cite{Banburski:2014cwa})
\be \label{eqn_det_id}
  \det\left( \one - \sum_i C_i |A_i \ket [B_i| \right) = 1 + \sum_i C_i [A_i|B_i\ket + \sum_{i<j} C_i C_j [A_i|A_j\ket [B_i|B_j\ket
\ee
Now let $b = \bra z_i,w_i \ket_{n-1}$ and $c = (z_i|w_i)_{n-1}$ and upon differentiating we find
\begin{align}
  \int \rd \mu(z_n) \int\rd \mu(w_n) \delta_{\text{BF}}(z_n,w_n) P(z_i,w_i) 
  &= -\left( \frac{\partial}{\partial \sigma} + 1 \right) \rd \mu_\sigma(z_n) P(z_i,w_i) \Big|_{\sigma = 1} \nonumber \\
  &= \frac{1}{2\pi i} \oint_{\gamma_0} \rd s \, e^{s + \frac{c}{s}} \frac{2\sigma s^2-sb- (\sigma^2s^2-\sigma s b+c)}{(\sigma^2s^2-\sigma sb+c)^2} \Big|_{\sigma = 1} \nonumber \\
  &= \frac{1}{2\pi i} \oint_{\gamma_0} \rd s \, e^{s + \frac{c}{s}} \frac{s^2-c}{(s^2-sb+c)^2}  \label{eqn_Loop_id_rough}
\end{align}
We know that the final result must be $e^b$. Indeed, this follows by a simple change of variable\footnote{We thank James Ryan for pointing this out.} $t = s + c/s$ 
\be \label{eqn_loop_id_final}
  \int_{\C^{2}} \rd \mu(z_n)\int_{\C^{2}}\rd \mu(w_n) \,\delta_{\text{BF}}(z_n,w_n) P(z_i,w_i)  
  = \frac{1}{2\pi i} \oint_{\gamma_0} \rd t \, \frac{e^{t} }{(t-b)^2}  = e^b
\ee
The mapping $t = s + c/s$ is a conformal transformation known as the Joukowsky transformation used to model airfoils. 

What is remarkable about this result is that the translation symmetry of BF theory is somehow captured in this change of variables from (\ref{eqn_Loop_id_rough}) to (\ref{eqn_loop_id_final}).  A geometric understanding of this transformation is left to future study \cite{Livine_to_come}, however we note that there has been interest in this direction \cite{Langvik:2016hxn}.  We will see next that the loop identity for the Wilson action can be derived via the same transformation.

\section{The Yang-Mills Loop Identity} \label{sec_YM_loop}

Similarly to how we inserted the dimension factor $2j+1$ into the trace of the loop identity, we would now like to insert the factor $(2j+1)I_{2j+1}(\beta)$ to match  (\ref{eqn_wilson_char_exp}).  To do this we use a trick as shown in \cite{Holtkamp:1981su}.
The idea is to express the reproducing kernel by the following identity
\be \label{eqn_prod_rep}
  \frac{\bra z|w\ket^{2j}}{(2j)!} = \frac{1}{2\pi i}\oint_{\gamma_0} \frac{\rd \tau}{\tau^{2j+1}}  e^{\tau \bra z|w\ket }
\ee
and hence we would like to perform the following sum
\be
  \sum_{j \in \N/2} (2j+1) I_{2j+1}(\beta) \tau^{-2j-1}
\ee
We can then use the fact that for non-negative powers of $\tau$ the integrals in (\ref{eqn_prod_rep}) vanish, so we can express this equivalently as
\be
  \sum_{j \in \N/2} (2j+1) I_{2j+1}(\beta) \left(\tau^{-2j-1} - \tau^{2j+1}\right) = \frac{\beta}{2}(\tau^{-1} - \tau) e^{ \frac{\beta}{2}\left(  \tau + \tau^{-1} \right)}
\ee
where the summation is performed by using (\ref{eqn_wilson_char_exp}) and the Weyl Character formula
\be
  \chi_{j}(g) = \frac{\tau^{-2j-1} - \tau^{2j+1}}{\tau^{-1} - \tau } \qquad \text{tr}(g) =\tau + \tau^{-1}  \qquad \tau \in U(1)
\ee
Therefore the Yang-Mills kernel has the closed form
\be \label{eqn_YM_phase}
  \delta_{\text{YM}}(z,w,\beta) \equiv \sum_{j \in \N/2} (2j+1) I_{2j+1}(\beta) \frac{\bra z|w\ket^{2j}}{(2j)!}
	=  \frac{1}{2 \pi i} \oint_{\gamma_{0}} \rd \tau \, \frac{\beta}{2}(\tau^{-1} - \tau) \, e^{ \frac{\beta}{2}\left(  \tau + \tau^{-1} \right) +  \tau \bra z|w\ket }
\ee
Replacing one reproducing kernel (\ref{eqn_rep_kern}) in a closed loop by (\ref{eqn_YM_phase}) will therefore introduce the heat kernel as in (\ref{eqn_wilson_char_exp}).  Let us now perform the loop identity calculation with the Yang-Mills phase (\ref{eqn_YM_phase}) included
\be
  \int_{\C^{2}} \rd \mu(z_n)\int_{\C^{2}} \rd \mu(w_n) \, \delta_{\text{YM}}(z_n,w_n,\beta) P(z_i,w_i) 
\ee
In the limit $\beta \rightarrow \infty$ we expect this to match the BF theory result, i.e. the RHS of (\ref{eqn_loop_id_beginning}) up to a factor $I_1(\beta)$ as in (\ref{eqn_heat_kernel_approx}).  Integrating out $z_n, w_n$ similarly to (\ref{eqn_loop_poles}) leads to 
\be
  \frac{(\beta/2)}{(2 \pi i)^2} \oint_{\gamma_{0}^2} \rd \tau \rd s \, (\tau^{-1} - \tau) \frac{e^{ \frac{\beta}{2}\left(  \tau + \tau^{-1} \right) +  s + cs^{-1} }}{s^2-s\tau b +\tau^2 c}
\ee
where again $b = \bra z_i,w_i \ket$ and $c = (z_i|w_i)_{n-1}$.  Shifting $s \rightarrow s\tau$ we can disentangle the polynomial in the denominator
\be
  \frac{(\beta/2)}{(2 \pi i)^2} \oint_{\gamma_{0}^2} \rd \tau \rd s \, (\tau^{-2} - 1) \frac{e^{ \tau\left(\frac{\beta}{2}+s\right) + \tau^{-1}\left(\frac{\beta}{2}+cs^{-1}\right) }}{s^2-s b + c}
\ee
Now using the Laurent expansion for the modified Bessel function $$e^{x\tau+y\tau^{-1}} = \sum_{m=-\infty}^{\infty} I_m\left(2\sqrt{xy}\right) \left(\sqrt{\frac{x}{y}} \,\tau\right)^m$$ we can perform the integral over $\tau$ to get
\be
  \frac{(\beta/2)}{2 \pi i} \oint_{\gamma_{0}}\frac{\rd s }{s^2-s b + c} I_{1}\left(2\sqrt{xy}\right)\left( \frac{x-y}{\sqrt{xy}} \right)
\ee
where $x = \frac{\beta}{2}+s$ and $y = \frac{\beta}{2}+cs^{-1}$.  Making the same change of variable $t = s +cs^{-1}$ as we did in Section \ref{sec_loop_id} we find
\be
  \frac{(\beta/2)}{2 \pi i} \oint_{\gamma_{0}}\frac{\rd t }{t-b} \frac{I_{1}\left(2\omega\right)}{\omega} = \frac{2(\beta/2)}{2 \pi i} \oint_{\gamma_{0}}\rd \omega  \frac{I_{1}\left(2\omega\right)}{(\omega-r)(\omega+r)}
\ee
where $\omega^2 = \left(\frac{\beta}{2}\right)^2 + \left(\frac{\beta}{2}\right) t +c$.  Finally change variable from $t$ to $\omega$ and integrate
\be
  \frac{2(\beta/2)}{2 \pi i} \oint_{\gamma_{0}}\rd \omega  \frac{I_{1}\left(2\omega\right)}{(\omega-r)(\omega+r)} = \frac{\beta }{2r} \, I_1(2r)
\ee
where $r^2 = (\beta/2)^2 + (\beta/2)b +c$.  The final result is therefore
\be \label{eqn_YM_loop_id_final}
  \int_{\C^{2}} \rd \mu(z_n)\int_{\C^{2}} \rd \mu(w_n) \, \delta_{\text{YM}}(z_n,w_n,\beta) P(z_i,w_i)  = \frac{\beta }{2r} \, I_1(2r)
\ee
Thus we have shown that the loop identity can be calculated using the contour integral representation of the projector with the non-trivial Yang-Mills kernel.  

We note that the integral representation of the modified Bessel function is the same form as the original projector (\ref{eqn_proj})
\be
   \frac{\beta }{2r} \, I_1(2r) = \frac{\beta}{2} \sum_{k=0}^{\infty} \frac{r^{2k}}{k!(k+1)!} = \frac{(\beta/2)}{2 \pi i} \oint_{\gamma_{0}}\frac{\rd s }{s^2} e^{s + s^{-1}\left( \left(\frac{\beta}{2}\right)^2 + \left(\frac{\beta}{2}\right) b + c \right)}
\ee
Note that the factor with $b$ is the trivial reproducing kernel on $n-1$ strands, which would be the result of the
BF loop identity, while the factor with $c$ is the projector on $n-1$ strands which is the non-triviality.  

Let us now check our guess for the asymptotic behaviour as $\beta \rightarrow \infty$.  Note that $I_m(r) \sim (2\pi r)^{-1/2} e^{r}$ and $r \sim \beta/2 + b/2$ therefore
\be
  \lim_{\beta \rightarrow \infty} \frac{\beta }{2r} \, I_1(2r) = (2\pi\beta)^{-\frac{1}{2}} \, e^{\beta + b}
\ee 
which agrees with the factor $I_1(\beta)$ in (\ref{eqn_heat_kernel_approx}).  
Finally we note that using the relation (\ref{eqn_YM_phase}) and (\ref{eqn_wilson_char_exp}) the Yang-Mills loop identity (\ref{eqn_YM_loop_id_final}) is in group variables
\be \label{eqn_YM_loop_group}
  \int_{\text{SU(2)}} \rd g \, e^{\sum_{i=1}^{n-1} \bra z_i | g | w_i \ket} e^{\frac{\beta}{2} \text{tr}(g)} = \frac{1}{r} \, I_1(2r)
\ee
As a consistency check we can use group integration techniques to perform the same calculation.  Indeed, using the formula (see the appendix of \cite{Freidel:2010tt} for proof)
\be
  \int_{\text{SU}(2)} \rd g \, e^{\text{tr}(gX)} = \sum_{J=0}^{\infty} \frac{\det(X)^J}{J!(J+1)!}
\ee
with $X = \beta/2 + \sum_{i=1}^{n-1} |w_i\ket\bra z_i|$ and using (\ref{eqn_det_id}) we have $\det(X) = r^2$ and so the LHS of (\ref{eqn_YM_loop_group}) matches the RHS.

\section{Reduction to the Grassmannian} \label{sec_closure}

To summarize up to this point, we can now construct the Yang-Mills partition function (\ref{eqn_YM_part_func}) using the projector (\ref{eqn_proj}) and the extra phase (\ref{eqn_YM_phase}).  Furthermore, using the square of the projector
\begin{align} \label{eqn_double_proj}
  \int_{\C^{2n}} \rd\mu(x_i) P(z_i,x_i) P(x_i,w_i) = \frac{1}{(2\pi i)^2}\oint_{\gamma_0}\oint_{\gamma_0} \frac{\rd s \rd t}{(st)^2} \int_{\C^{2n}} \rd\mu(x_i) e^{s+t + \frac{1}{s}(z_i|x_i) +  \frac{1}{t}(x_i|w_i)} 
\end{align}
the partition function conveniently factorizes at the vertices as the vertex amplitudes (\ref{eqn_proj_cont}).  We would now like to perform the reduction by $\C^{2n}/\text{GL}(2,\C) \cong \text{Gr}(2,n)$ on the spinors $\{x_i\}$, which we note does not affect the form of the vertex amplitudes.

As shown in \cite{Fujii:1995ds} and \cite{Freidel:2010tt} this can be achieved by imposing the so called closure constraints on the $x_i$:
\be \label{eqn_closure_const}
  \sum_i |x_i \ket \bra x_i | = 1
\ee
where $1$ is the $2 \times 2$ identity matrix.  We do this by a Fadeev-Popov like procedure using the following integral evaluation over $\text{GL}(2,\C)$
\be \label{eqn_edge_constraints}
  \int_{\text{GL}(2,\C)} \frac{\rd^8 g}{|\det(g)|^4} \delta^{(4)}\left( \sum_i g^{-1} |x_i \ket \bra x_i | (g^{-1})^{\dagger} - 1 \right) = \frac{\text{Vol}(\text{U}(2)) }{2}
\ee
the details of which are given in Appendix \ref{app_FP_det}.  Inserting this equality into (\ref{eqn_double_proj}) and making a change of variable $|x_i\ket = g |y_i\ket$
\begin{align}
	& \int_{\C^{2n}} \rd\mu(x_i) P(z_i,x_i) P(x_i,w_i) \int\limits_{\text{GL}(2,\C)} \frac{\rd^8 g}{|\det(g)|^4} \frac{\delta^{(4)}\left( \sum_i g^{-1} |x_i \ket \bra x_i | (g^{-1})^{\dagger}  - 1 \right)}{\frac{1}{2}\text{Vol}(U(2)) } \\
	&= \int\limits_{\text{GL}(2,\C)} \frac{\rd^8 g e^{-\text{tr}(gg^\dagger)}}{|\det(g)|^{4-2n}} \oint_{\gamma_0}  \oint_{\gamma_0} \frac{\rd s \, \rd t}{s^2 t^2}  \, \frac{e^{s + t}}{(2\pi i)^2} \int_{\C^{2n}} \prod_i \frac{\rd^4 y_i}{\pi^2}  e^{\frac{\det(g)}{s}(z_i|y_i) +  \frac{\det(g^\dagger)}{t}(y_i|w_i)} \frac{\delta^{(4)}\left( \sum_i |y_i \ket \bra y_i |  - 1  \right)}{\frac{1}{2}\text{Vol}(\text{U}(2)) } \nn 
\end{align}
Now making a change of variables $s' = s/\det(g)$ and $t' = t/\det(g^\dagger)$ this becomes
\begin{align}
	\frac{1}{(2\pi i)^2} \oint_{\gamma_0}  \oint_{\gamma_0} \frac{\rd s' \, \rd t'}{s'^2 t'^2}  \int_{\C^{2n}} \frac{\rd \Omega(y_i)}{\pi^{2n}} e^{\frac{1}{s'}(z_i|y_i) +  \frac{1}{t'}(y_i|w_i)} \int\limits_{\text{GL}(2,\C)} \frac{\rd^8 g}{|\det(g)|^{6-2n}}   e^{-\text{tr}(gg^\dagger) + s'\det(g) +  t'\det(g^\dagger)} 
\end{align}
where we define the measure
\be \label{eqn_measure_grass} 
\rd \Omega(y_i) \equiv \frac{2}{\text{Vol}(\text{U}(2)) } \delta^{(4)}\left( \sum_i |y_i \ket \bra y_i |  - 1 \right) \prod_i \rd^4 y_i.
\ee 
Let us immediately drop the primes on $s', t'$ and change $y_i$ back to $x_i$.  We can perform the integral over $g$ in the following way.  Let us write $g$ in terms of a pair of independent spinors $u, v$ in the standard orthonormal basis $\bra 0 | = (1,0)$, $[0| = (0,1)$ 
$$
  g = |u\ket\bra 0 | + |v \ket [0|
$$
so $\det(g) = [u|v\ket$ and $\text{tr}(gg^\dagger) = \bra u | u \ket + \bra v | v \ket$.  Then the integral becomes
\begin{align} \label{eqn_int_derivatives}
  \int\limits_{\text{GL}(2,\C)} \rd^8 g \frac{e^{- \text{tr}(gg^\dagger) + s\det(g) +  t\det(g^\dagger)}}{|\det(g)|^{6-2n}}    
	&= \frac{\partial^{2n-6}}{\partial \sigma^{n-3}\partial \tau^{n-3}}\int_{\C^4} \rd^4 u \, \rd^4 v e^{-\bra u | u \ket - \bra v | v \ket + (s+\sigma)[u|v\ket + (t+\tau)\bra v | u]} \bigg|_{\sigma=\tau=0} \nn \\
	&= \frac{\partial^{2n-6}}{\partial \sigma^{n-3}\partial \tau^{n-3}}\frac{\pi^4}{ \left(1-(s+\sigma)(t+\tau)\right)^2} \bigg|_{\sigma=\tau=0} 
\end{align}
For $n=3,4$ this gives respectively
$$
  \pi^4\frac{1}{(1-st)^2} \qquad(2\pi^4)\frac{1+2st}{(1-st)^4}
$$
We can solve for this function of $s,t$ explicitly for any $n$ in the following way.  First we expand the exponentials and perform the integrals
\begin{align} \label{eqn_GL2C_F}
  \int\limits_{\text{GL}(2,\C)} \rd^8 g \frac{e^{- \text{tr}(gg^\dagger) + s\det(g) +  t\det(g^\dagger)}}{|\det(g)|^{6-2n}}    
	&= \sum_{J,J'=0}^{\infty} \int_{\C^4} \rd^4 u \, \rd^4 v e^{-\bra u | u \ket - \bra v | v \ket} \frac{[u|v\ket^{J+n-3}}{J!} \frac{\bra v | u]^{J'+n-3}}{J'!} s^J t^{J'} \nn \\
	&= \pi^2 \sum_{J=0}^{\infty} \int_{\C^2} \rd^4 u  e^{-\bra u | u \ket } \frac{[u|u]^{J+n-2}}{J!J!} (J+n-3)! s^Jt^J \nn \\
	&= \pi^4 \sum_{J=0}^{\infty} \frac{(J+n-2)!(J+n-3)!}{J!J!} (st)^J \nn \\
	&= \pi^4 (n-2)!(n-3)!{}_{2}F_1(n-1,n-2;1;st)
\end{align}
and the sum converges so long as $|st| < 1$ which we can safely assume.  Lets put this all together now.  First, we know from \cite{Freidel:2010tt} that the integral over the measure $\rd \Omega$ is related to the volume of the Grassmannian (see Appendix \ref{app_vol_grass}).  Therefore let us define the normalized measure
\be \label{vol_grassmannian}
 \rd \hat{\Omega}(x_i) = \frac{(n-1)!(n-2)!}{\pi^{2(n-2)}} \rd \Omega(x_i) \qquad \text{where} \qquad \int_{\C^{2n}} \rd \hat{\Omega}(x_i) = 1 
\ee
Second let us define 
\be
  K_n(st) \equiv \frac{{}_{2}F_1(n-1,n-2;1;st) }{(n-1)(n-2)(st)^2} 
\ee
Thus the projector becomes
\begin{align}
	P(z_i,w_i) = \oint_{{\gamma_0}} \oint_{{\gamma_0}} \frac{\rd s \, \rd t}{(2\pi i)^2} K_n(st)  \int_{\C^{2n}} \rd \hat{\Omega}(x_i) e^{\frac{1}{s}(z_i|x_i) +  \frac{1}{t}(x_i|w_i)}   
\end{align}
Finally, let us include factors of $(x_i|x_i) = \det(\sum_i |x_i\ket \bra x_i|)$ which is equal to one because of the delta function.  In this way we can make the integrand of the projector manifestly $\text{GL}(2,\C)$ invariant  
\begin{align} \label{eqn_projector_with_kernel}
	P(z_i,w_i) = \frac{1}{(2\pi i)^2}\int_{\C^{2n}} \rd \hat{\Omega}(x_i) \oint_{{\gamma_0}} \oint_{{\gamma_0}} \frac{\rd s \, \rd t}{(x_i|x_i)} K_n\left(\frac{st}{(x_i|x_i)}\right) e^{\frac{1}{s}(z_i|x_i) +  \frac{1}{t}(x_i|w_i)}   
\end{align}
The advantage of this formula over the one given in \cite{Freidel:2010tt} is that it is Gaussian while $z_i$ and $w_i$ are still factorized as in (\ref{eqn_double_proj}).  This will allow us to contract these projectors at vertices in a nice way, i.e. as the vertex amplitudes defined by cycles as we did in Section \ref{sec_vertex_amp}.

\subsection{Consistency Check}

As a consistency check if we set the external spinors to zero $z_i = w_i = 0$ in (\ref{eqn_projector_with_kernel}) we must have
\begin{align}
	\oint_{\gamma_0}  \oint_{\gamma_0} \frac{\rd s \, \rd t}{(2\pi i)^2} K_n(st) = 1
\end{align}
since $\rd\hat{\Omega}$ is a normalized measure.  Let us check this
\begin{align}
	\oint_{\gamma_0}  \oint_{\gamma_0} \frac{\rd s \, \rd t}{(2\pi i)^2} \frac{{}_{2}F_1(n-1,n-2;1;st) }{(n-1)(n-2)(st)^2}
	&= \sum_{J=0}^{\infty} \frac{(J+n-2)!(J+n-3)!}{J!J!(n-1)!(n-2)!} \oint_{\gamma_0}  \oint_{\gamma_0} \frac{\rd s \, \rd t}{(2\pi i)^2} (st)^{J-2} \nn \\ 
	&= \frac{(J+n-2)!(J+n-3)!}{J!J!(n-1)!(n-2)!} \bigg|_{J=1} 
	= 1
\end{align}
Now let us compare (\ref{eqn_projector_with_kernel}) with the formula given in \cite{Freidel:2010tt}.  Let us denote $P(z_i,w_i) = \sum_J P_J(z_i,w_i)$ so $P_J(z_i,w_i) = (z_i|w_i)^J/J!(J+1)!$.  Then the formula in \cite{Freidel:2010tt} reads
\be \label{eqn_UN_coh_formula}
  \int_{\C^{2n}} \rd\mu(x_i) P_J(z_i,x_i) P_{J}(x_i,w_i)
  = \frac{D_{n,J}}{J!(J+1)!} \int_{Gr(2,n)} \rd \hat{\Omega}(x_i) \frac{(z_i|x_i)^J (x_i|w_i)^J}{(x_i|x_i)^J}
\ee
where $D_{n,J}$ is the dimension of the U($n$) representation of highest weight $[J,J,0,0,\cdots,0]$ given by
\be
  D_{n,J} = \frac{(J+n-1)!(J+n-2)!}{J!(J+1)!(n-1)!(n-2)!}, 
\ee
To compare this with our formula let us expand the exponentials in (\ref{eqn_projector_with_kernel}) and integrate term by term
\begin{align} 
	&\sum_{J,J'=0}^{\infty} \int_{\C^{2n}} \rd\mu(x_i) P_J(z_i,x_i) P_{J'}(x_i,w_i)  
	=  \oint_{\gamma_0}  \oint_{\gamma_0}\frac{\rd s \, \rd t}{(x_i|x_i)}  \int_{\C^{2n}} \rd \hat{\Omega}(x_i) \frac{e^{\frac{1}{s}(z_i|x_i) +  \frac{1}{t}(x_i|w_i)}}{(2\pi i)^2} K_n\left(\frac{st}{(x_i|x_i)}\right) \nn \\
	&= \oint_{\gamma_0}  \oint_{\gamma_0} \frac{\rd s \, \rd t}{(2\pi i)^2}  \int_{\C^{2n}} \rd \hat{\Omega}(x_i)  \sum_{J,J'=0}^{\infty} \frac{(z_i|x_i)^J}{J!} \frac{(x_i|w_i)^{J'}}{J'!}  \sum_{k=0}^{\infty} \frac{(k+n-2)!(k+n-3)!}{k!k!(n-1)!(n-2)!} \frac{s^{k-J-2} t^{k-J'-2}}{(x_i|x_i)^{k-1}}\nn \\
	&= \int_{\C^{2n}} \rd \hat{\Omega}(x_i)  \sum_{J=0}^{\infty} \frac{(z_i|x_i)^J(x_i|w_i)^{J}}{(x_i|x_i)^{J}} \frac{(J+n-1)!(J+n-2)!}{J!J!(J+1)!(J+1)!(n-1)!(n-2)!} 
\end{align}
which indeed matches (\ref{eqn_UN_coh_formula}).

\subsection{The Standard Gauge}

Another way of fixing the $\text{GL}(2,\C)$ invariance in (\ref{eqn_projector_with_kernel}) is to fix two of the $n$ spinors, say $x_1$ and $x_2$ to an arbitrary pair of linearly independent spinors $y_1$, $y_2$. The matrix $g \in \text{GL}(2,\C)$ which maps $|y_1\ket, |y_2\ket$ to $|x_1\ket, |x_2\ket$ is by inspection
\be
  g = \frac{|x_2\ket[y_1| - |x_1\ket[y_2|}{[y_1|y_2\ket} \qquad \det(g) = \frac{[x_1|x_2\ket}{[y_1|y_2 \ket}
\ee
which is well defined so long as $y_1$ and $y_2$ are linearly independent.  Let us denote the integrand of (\ref{eqn_projector_with_kernel}) by $F(x_i)$, which as we noted is $\text{GL}(2,\C)$ invariant.  Define $y_i$ for $i=3,...,n$ by
\be
  g|y_i\ket = |x_i\ket \qquad i=1,...,n
\ee
where $i=1,2$ are by construction.  By $\text{GL}(2,\C)$ invariance $F(x_i) = F(y_i)$ while the integration measure (\ref{vol_grassmannian}) neglecting the normalization factor becomes
\begin{align}
 \delta^{(4)}\left( \sum_{i=1}^{n} |x_i \ket \bra x_i |  - 1 \right) \prod_{i=1}^{n} \rd^4 x_i 
 =
 \delta^{(4)}\left( \sum_{i=1}^{n} g |y_i \ket \bra y_i | g^\dagger  - 1 \right)  \frac{\rd^8 g}{|\det(g)|^{4-2n}} \, |[y_1|y_2\ket|^4 \, \prod_{i=3}^{n} \rd^4 y_i
\end{align}
where we used
\be
  \rd^4 x_1 \, \rd^4 x_2 = |[y_1|y_2\ket|^{4} \, \rd^8 g  \qquad \text{and} \qquad  \rd^4 x_i = |\det(g)|^2 \, \rd^4 y_i \quad i=3,...,n
\ee
We would now like to perform the integration over $\rd^8 g$:  
\be
  \int_{\text{GL}(2,\C)} \rd^8 g \, |\det(g)|^{2n-4 } \, \delta^{(4)}\left( \sum_{i=1}^{n} g |y_i \ket \bra y_i | g^\dagger  - 1 \right)  
\ee
Since the measure is $U(2)$ invariant, we can assume $\sum_i |y_i\ket \bra y_i|$ is diagonal and take $\lambda_1, \lambda_2$ to be its eigenvalues.  Expressing $g$ as a pair of spinors $u, v$ in a basis $\bra 0 | 0 \ket = [0|0] = 1$, $[0|0\ket=0$
$$
  g = |u\ket\bra 0 | + |v \ket [0| \qquad  \det(g) = [u|v\ket
$$
the integral becomes
\be
  \int_{\C^4} \rd^4 u \rd^4 v \, |[u|v\ket|^{2n-4} \, \delta^{(4)}\left(  \lambda_1|u\ket\bra u| + \lambda_2|v\ket\bra v| - 1 \right)
\ee
Rescaling $u$ and $v$ we can eliminate $\lambda_1$ and $\lambda_2$ from the integrand giving an overall factor $(\lambda_1 \lambda_2)^{-n} = \det(\sum_{i} |y_i\ket \bra y_i|)^{-n}$.  Now defining 
\be
  \vec{U} = \bra u | \vec{\sigma} | u \ket \qquad \vec{V} = \bra v | \vec{\sigma} | v \ket
\ee
we can express $|u\ket\bra u| = (|\vec{U}|+\vec{U}\cdot \vec{\sigma})/2$ and $|u][ u| = (|\vec{U}|-\vec{U}\cdot \vec{\sigma})/2$ so
\be
  [v|u\ket\bra u | v] = \frac{|\vec{U}| |\vec{V}|-\vec{U} \cdot \vec{V}}{2}
\ee
Expressing the integrations over 3-vectors as in (\ref{eqn_spinor_spherical}) we have
\be
  \frac{1}{(\lambda_1 \lambda_2)^{n}} \left(\frac{\pi}{4}\right)^2 \int \frac{\rd^3 \vec{U} \rd^3 \vec{V}}{|\vec{U}| |\vec{V}|} \left(\frac{|\vec{U}| |\vec{V}|-\vec{U} \cdot \vec{V}}{2}\right)^{n-2} \delta^{(1)}\left(  |\vec{U}| + |\vec{V}| - 1 \right) \delta^{(3)}\left( \vec{U} + \vec{V} \right)
\ee
Performing the integration over $\vec{V}$ and then $\vec{U}$
\be
  \frac{8}{(\lambda_1 \lambda_2)^{n}} \left(\frac{\pi}{4}\right)^2 \int \rd^3 \vec{U} \, |\vec{U}|^{2n-6} \, \delta\left(  |\vec{U}|  - 1 \right) 
	= \frac{\text{Vol}(U(2))}{2(\lambda_1 \lambda_2)^{n}}
\ee
where we identified $\text{Vol}(\text{U}(2)) = 4\pi^3$.  Thus in conclusion
\be
  \int_{\C^{2n}}\, \delta^{(4)}\left( \sum_{i=1}^{n} |x_i \ket \bra x_i |  - 1 \right) \prod_{i=1}^{n} \rd^{4} x_i 
  =
  \frac{\text{Vol}(U(2))}{2} \int_{\C^{2n-4}} \, \frac{|[y_1|y_2\ket|^{4} }{\det(\sum_{i=1}^{n} |y_i\ket \bra y_i|)^{n}} \, \prod_{i=3}^{n} \rd^{4} y_i 
\ee
Using the definitions (\ref{eqn_measure_grass}, \ref{vol_grassmannian}) of the measure $\rd \hat{\Omega}$ we get
\be
  \rd \hat{\Omega}(x_i) =  \frac{|[x_1|x_2\ket|^{4} \prod_{i=1}^{n} \rd^{4} x_i}{\text{vol}(\text{Gr}(2,n))(x_i|x_i)^n} \delta^{(4)}(x_1 - y_1) \delta^{(4)}(x_2 - y_2)
\ee
where $\text{vol}(\text{Gr}(2,n)) = \pi^{2(n-2)}/(n-1)!(n-2)!$.  Putting this into (\ref{eqn_projector_with_kernel}) the projector becomes
\begin{align} \label{eqn_gauge_fix_prop}
	P(z_i,w_i) = \frac{1}{(2\pi i)^2} \int_{\C^{2n-4}}  \frac{|[y_1|y_2\ket|^{4} \prod_{i=3}^{n} \rd^{4} y_i}{\text{vol}(\text{Gr}(2,n))(y_i|y_i)^n} \,  \oint_{{\gamma_0}} \oint_{{\gamma_0}} \frac{\rd s \, \rd t}{(y_i|y_i)} K_n\left(\frac{st}{(y_i|y_i)}\right) e^{\frac{1}{s}(z_i|y_i) +  \frac{1}{t}(y_i|w_i)}   
\end{align}
with $y_1$ and $y_2$ chosen constant and linearly independent but otherwise arbitrarily.

\subsection{Consistency Check}

As a check if we let the external spinors vanish $z_i = w_i = 0$ in (\ref{eqn_gauge_fix_prop}) then we get the relation
\be
   \int_{\C^{2n-4}}  \frac{|[y_1|y_2\ket|^{4} \prod_{i=3}^{n} \rd^{4} y_i}{(y_i|y_i)^n} = \text{vol}(\text{Gr}(2,n))
\ee
Now consider the case $n=3$ with the gauge fixing $|y_1\ket = |0\ket$, $|y_2\ket = |0]$.  Then $[y_1|y_2\ket =1$ and
\be
  (y_i|y_i) = \det\left( 1 + |y_3\ket\bra y_3|\right) = 1 + \bra y_3 | y_3 \ket
\ee
Hence we have
\begin{align} \label{eqn_stand_gauge_prop}
	\int_{\C^{2}} \frac{\rd^{4} y_3}{(1 + \bra y_3 | y_3 \ket)^3} = \frac{\pi}{4} \int_{\R^3} \frac{\rd^3 \vec{Y}}{|\vec{Y}|} \frac{1}{(1+|\vec{Y}|)^3} = \frac{\pi^2}{2}
\end{align}
where in the first step we used (\ref{eqn_spinor_spherical}) with $\vec{Y} \equiv \bra y_3 | \vec{\sigma}|y_3\ket$ and in the second step we performed the integral.  This agrees with the formula for $\text{vol}(\text{Gr}(2,3))$. 

\subsection{Cross-Ratios}

In this section we will show how the standard gauge is related to the coordinates introduced in \cite{Freidel:2009nu} in terms of cross ratios.  The representation of SU(2) chosen there is on $\C P 1$ rather than $\C^2$ but they are clearly related by choosing homogeneous coordinates on $\C P 1$.  Explicitly, $(z_1, z_2) \sim z_1/z_2 \in \CP1$.  

In \cite{Freidel:2009nu} $n$ copies of $\C P1$ label an SU(2) intertwiner.  A SL(2,$\C$) transformation can then be used to fix, three of the coordinates to three arbitrary points which they take to be 0,1, and $\infty$.  In homogeneous coordinates, 0 and $\infty$ correspond to $\lambda|0]$ and $\mu|0\ket$ respectively for arbitrary $\lambda, \mu \in \C$.

A third spinor can be fixed to $1 \sim |0\ket + |0] \equiv |1\ket$ by a further GL(2,$\C$) transformation.  Indeed, let $|y_1\ket = |0\ket$, $|y_2\ket = |0]$ in (\ref{eqn_gauge_fix_prop}) and define
\be
  g = \lambda^{-1} |0\ket \bra 0 | + \mu^{-1} |0][ 0 |
\ee
where $|y_3\ket = (\lambda,\mu)^t$.  Then $g|0\ket =\lambda^{-1}|0\ket$, $g|0] = \mu^{-1}|0]$, and $g|y_3\ket = |1\ket$.  Performing the change of variables $y_i \rightarrow g y_i$ for $i=3,...,n$ in (\ref{eqn_gauge_fix_prop}) we get
\begin{align} 
	P(z_i,w_i) = \frac{1}{(2\pi i)^2} \int_{\C^{2n-4}} \frac{|\mu\lambda|^4 \rd^2 \lambda \rd^2 \mu \prod_{i=4}^{n} \rd^{4} y_i}{\text{vol}(\text{Gr}(2,n))(y_i|y_i)^n} \,  \oint_{{\gamma_0}} \oint_{{\gamma_0}} \frac{\rd s \, \rd t}{(y_i|y_i)} K_n\left(\frac{st}{(y_i|y_i)}\right) e^{\frac{1}{s}(z_i|y_i) +  \frac{1}{t}(y_i|w_i)}   
\end{align}
where now $y_1 = \lambda|0\ket$, $y_2 = \mu|0]$, and $y_3 = |1\ket$.  The remaining variables are 
\be
  g|y_i\ket = \frac{[ y_2|y_i\ket}{[ y_2 |y_3 \ket} |0\ket + \frac{[ y_1|y_i\ket}{[ y_1 |y_3 \ket} |0] \qquad i=4,...,n
\ee
and hence are the cross ratios in inhomogeneous coordinates.  This is thus the analogous formula for the measure given in \cite{Freidel:2009nu}.  
\section{On Simplicity Constraints}

As mentioned in the introduction the main interest in spin foam models is as state sum models for quantum gravity.  Such models are based on the Plebanski action for General Relativity which is the continuum four dimensional BF action with an added set of constraints called simplicity constraints \cite{Plebanski}.  The simplicity constraints force the two-form B field to be simple, i.e. the wedge product of two real one-forms, which breaks the topological gauge symmetries of BF theory and gives rise to the local degrees of freedom of General Relativity  \cite{DePietri:1998mb} . 

Since the continuum Plebanski path integral is ill-defined, the strategy is to use the spin foam representation on a discretization and to apply analogous constraints.  Here we will briefly discuss two variants of the so called Holomorphic Simplicity Constraints \cite{Dupuis:2011fz} which are most suitable for the framework we have introduced in this paper.  We consider four dimensional Euclidean spacetime, and hence the gauge group $\text{Spin}(4) \cong \text{SU}(2)\times \text{SU}(2)$.

Due to this isomorphism, the partition function of $\text{Spin}(4)$ Yang-Mills is simply the product
\be
  Z_{\text{YM}}^{\text{Spin}(4)}(\beta) = Z_{\text{YM}}^{\text{SU}(2)}(\beta) Z_{\text{YM}}^{\text{SU}(2)}(\beta)
\ee
and the $\text{Spin}(4)$ projector is the product of two SU(2) projectors
\be \label{eqn_spin_4_proj}
  P^{\text{Spin}(4)}\left(\{z_i,z'_i\},\{w_i,w'_i\}\right) =
  P^{\text{SU}(2)}(z_i,w_i)P^{\text{SU}(2)}(z'_i,w'_i)
\ee
where a prime distinguishes the left and right copies of SU(2).  For more details see \cite{Banburski:2014cwa}.

The Holomorphic simplicity constraints proposed in \cite{Dupuis:2011fz} are applied to the squared projector
\be
  P\left(\{z_i,z'_i\},\{w_i,w'_i\}\right) = \int \rd \mu(x_i) P(z_i,x_i)P(x_i,w_i) \int \rd \mu(x'_i) P(z'_i,x'_i)P(x'_i,w'_i)
\ee
by imposing the constraint $x_i = \rho x'_i$ for $\rho \in \R$.  This is essentially a constraint on vertex amplitudes given in section \ref{sec_vertex_amp}.  

There is a subtly with these constraints, namely that their validity requires the closure constraints (\ref{eqn_closure_const}) to be satisfied.  This is not a problem in the asymptotic limit, however one can also use the projector on the Grassmannian (\ref{eqn_projector_with_kernel}) to ensure that the closure constraints are enforced strongly.  Doing so we see that $\rho$ is a relative scale factor for the left and right polyhedra.  The significance of $\rho$ within these constraints with respect to scale invariance is left as a topic for future study.

A simpler version of the Holomorphic simplicity constraints was investigated in \cite{Banburski:2014cwa}.  There, instead of using the squared projector (and hence vertex amplitudes) one uses the projector $P(z_i,w_i)P(z'_i,w'_i)$ directly and sets $z_i = \rho z'_i$ and $w_i = \rho w'_i$.  Again closure must be assumed, but is valid in the asymptotic limit.  In the contour integral representation this constrained projector is
\be \label{eqn_const_proj}
 P(z_i,w_i)P(\rho z_i, \rho w_i) = \frac{1}{(2\pi i)^2}\oint_{\gamma^{2}_{0}} \frac{\rd s \, \rd t}{s^2 \, t^2} \, e^{s + t + \left(\frac{1}{s} + \frac{\rho}{t}\right)(z_i|w_i)}
\ee
In \cite{Banburski:2014cwa} the loop identity for this constrained projector was computed in the spin representation, leading to some complicated but exact expressions.  Here a computation of the loop identity for (\ref{eqn_const_proj}) might have a more compact expression in the contour representation.  However, the conformal transformation leading to the BF and Yang-Mills loop identities does not seem applicable in this case.  

A geometrical understanding of the loop identity is needed to shed light on the implications of the simplicity constraints in this framework.

\section{Conclusion}

We've shown how to represent SU(2) Lattice gauge theory in arbitrary dimensions, on a general 2-complex, as a path integral over a product of vertex amplitudes.  These vertex amplitudes are given by the spin network generating function of the boundary graph labelled by elements of the Grassmannian.  Explicitly the partition function (\ref{eqn_YM_part_func}) takes the form
\be
  Z_{\text{YM}}(\beta) = \oint\limits_{\gamma_{0}^{F}} \frac{\rd \tau_f}{2\pi i}  K_{\text{YM}}(\tau_f,\beta) \oint\limits_{\gamma_{0}^{2E}} \frac{\rd s_e \rd t_e}{(2\pi i)^2} K_{n_e}(s_et_e)  \int\limits_{\text{Gr}(2,n_e)^E} \rd \hat{\Omega}(x^{e}_{f}) \, \prod_v A_{\Gamma_v}(x^{e}_f,s_e,t_e,\tau_f)
\ee
where $\gamma_0$ is a contour which encircles the origin in a counter-clockwise manner.  The kernel $ K_{\text{YM}}(\tau_f,\beta) \equiv \frac{\beta}{2}(\tau^{-1}_f - \tau_f)e^{ \frac{\beta}{2}\left(  \tau_f + \tau^{-1}_f \right)} $ is responsible for the Yang-Mills regularization which approaches BF theory in the $\beta \rightarrow \infty$ limit.  The vertex amplitudes $A_{\Gamma_v}(x^{e}_f,s_e,t_e,\tau_f)$ are given by the spin network generating functional of the boundary graph $\Gamma_v$ dual to $v$, which can be expanded in cycles of $\Gamma_v$ as explained in Section \ref{sec_vertex_amp}.  Also $\tau_f$ appears in $A_{\Gamma_v}$ for exactly one vertex $v$ of a face $f$.  For clarity, as an explicit example, a cubic lattice in three dimensions would have $n_e = 4$ as there are four square plaquettes per edge, and $\Gamma_v$ would be an octahedral graph where the six nodes are dual to the six sides of a cube.  

While this may not look like an improvement, the contour variables appear as simple poles in the vertex amplitudes.  Moreover we are most interested in studying the geometrical interpretation of the Grassmannian and contour variables with respect to the dynamics, and in particular with respect to conformal transformations.

These Grassmannian elements can be interpreted as framed polyhedra embedded in $\R^3$ as given by Minkowski's theorem \cite{Livine:2013tsa}.  Indeed, the closure constraint (\ref{eqn_closure_const}) implies
\be
  \sum_{i=1}^{n} \vec{V}_i = 0
\ee
where $\vec{V}_i = \bra x_i | \vec{\sigma}|x_i\ket$ which is a sufficient condition for the existence of a convex polyhedron having $n$ faces of area $\propto |V_i|^2 = \bra x_i | x_i \ket$.  However, there is an extra condition in (\ref{eqn_closure_const}) namely
\be
  \sum_{i=1}^{n} |V_i|^2 = 2
\ee
and hence the polyhedron is of fixed total area.  More generally, the GL(2,$\C$) invariance implies that these polyhedra are invariant under local scale transformations.  It would be interesting to understand the canonical action of the conformal group on the Grassmannian with regards to this polyhedral interpretation.  This is left as a topic for future investigation \cite{Livine_to_come}.

Furthermore, our ability to perform the reduction $\C^{2n}/\text{GL}(2,\C) \cong \text{Gr}(2,n)$ in closed form was aided by the introduction of the contour integrals of the variables $s,t$.  The geometrical interpretation of these variables and the conformal transformations leading to the loop identity is also left for future study.

We note that the loop identity derived for Yang-Mills theory has a nice closed form.  This allows for an exact calculation of the Pachner moves in Lattice Yang-Mills theory similarly to what was done in the context of Riemannian Spin Foam models \cite{Banburski:2014cwa}.  Moreover, using the orthogonality of the modified Bessel functions one could derive interesting identities such as recurrence relations between vertex amplitudes for fixed spins \cite{Bonzom:2009zd}.

Finally, the closure constraints in the notation of section \ref{sec_vertex_amp}
\be \label{eqn_close_conc}
  \sum_{j} | x^{i}_{j} \ket \bra x^{i}_{j} | -1 = 0
\ee
were used to perform the reduction to the Grassmannian in Section  \ref{sec_closure}.  These constraints Poisson commute with another set of constraints known as matching constraints
\be \label{eqn_matching}
  \bra x^{i}_{j} | x^{i}_{j} \ket - \bra x^{j}_{i} | x^{j}_{i} \ket = 0
\ee
and form a first class constraint system.  These constraints generate U(1) gauge transformations for each face of the polyhedron, rotating the frame on each face.  Thus a further symplectic reduction by the torus of diagonal U($n$) matrices is possible resulting in the Kapovich and Millson moduli space \cite{kapmil}.  Discrete geometries satisfying the constraints (\ref{eqn_close_conc}) and (\ref{eqn_matching}) and  are known as closed twisted geometries \cite{Freidel:2010aq} and are a generalization of Regge geometries.  In the case of three dimensions closed twisted geometries and Regge geometries are equivalent and hence the integrals over $x^{i}_{j}$ should localize completely.

\acknowledgments

We thank Etera Livine, James Ryan, Klaas Landsman, and Renate Loll for useful discussions.  This work is part of the research program of the Foundation for Fundamental Research on Matter (FOM), All which is part of the Netherlands Organization for Scientific Research (NWO).

\appendix

\section{Change of Coordinates}
Parametrize a spinor in Cayley-Klein coordinates
\be
  |x \ket = \bpm x_0 \\ x_1\epm = \bpm r \cos \frac{\theta}{2} e^{-i(\phi+\psi)/2} \\ r \sin \frac{\theta}{2} e^{i(\phi-\psi)/2} \epm
\ee
then the measure is
\be
  \rd^4 x = \rd^2 x_0 \, \rd^2 x_1= \frac{1}{4} r^3 \sin \theta \rd r \, \rd \theta \, \rd \phi \, \rd \psi
\ee
where $(r,\theta,\phi,\psi) \in [0,\infty) \times [0,\pi) \times [0, 2\pi) \times [0,2\pi)$.  
Therefore, for a $U(1)$ invariant function $f(x)$
\begin{align}
  \int_{\R^4} \rd^4 x \, f(x)
  =  \frac{1}{4} \int_{\R^+ \times S^3} r^3 \sin \theta \rd r \, \rd \theta \, \rd \phi \, \rd \psi \,  f(x) 
  =  \frac{\pi}{2} \int_{0}^{\infty} \rd r \, r^3 \int_{0}^{\pi} \rd \theta \, \sin \theta \int_{0}^{2\pi} \rd \phi \,  f(r,\theta,\phi)
\end{align}
Defining $\vec{X} = \bra x | \vec{\sigma} | x \ket$ then $r^2 = |\vec{X}|$ and 
\begin{align}
  \bra x | \sigma_1 | x \ket = r^2 \sin \theta \cos \phi \quad \bra x | \sigma_2 | x \ket = r^2 \sin \theta \sin \phi \quad \bra x | \sigma_3 | x \ket = r^2 \cos \theta 
\end{align}
so
\be \label{eqn_spinor_spherical}
  \int_{\C^2}\rd^4 x \, f(x)  =  \frac{\pi}{4} \int_{0}^{\infty} \rd |\vec{X}| \, |\vec{X}| \int_{0}^{\pi} \rd \theta \, \sin \theta \int_{0}^{2\pi} \rd \phi \,  f(r,\theta,\phi) = \frac{\pi}{4}\int_{\R^3} \frac{\rd^3 \vec{X}}{|\vec{X}|} f(\vec{X})
\ee

\section{Fadeev-Popov Determinant}
\label{app_FP_det}

We want to perform the integration
\be
  \int\limits_{\text{GL}(2,\C)} \frac{\rd^8 g}{|\det(g)|^4} \delta^{(4)}\left( \sum_i g^{-1} |x_i \ket \bra x_i | (g^{-1})^{\dagger} - 1 \right)
\ee
Since the measure is U(2) invariant, we can assume $\sum_i |x_i\ket \bra x_i|$ is diagonal and take $\lambda_1, \lambda_2$ to be its eigenvalues.  Expressing $g$ as a pair of spinors $u, \textbf{}v$ in a basis $\bra 0 | 0 \ket = [0|0] = 1$, $[0|0\ket=0$
$$
  g = |0][ u | - |0 \ket [v| \qquad g^{-1} = \frac{|u\ket\bra 0 | + |v \ket [0|}{[u|v\ket}
$$
the determinant is $\det(g) = [u|v\ket$ and the integrand becomes
\be
  \int_{\C^4} \frac{\rd^4 u \, \rd^4 v}{|[u|v\ket|^4} \, \delta^{(4)}\left(  \frac{\lambda_1|u\ket\bra u| + \lambda_2|v\ket\bra v|}{|[u|v\ket|^2} - 1 \right)
\ee
Factoring out the $|[u|v\ket|^2$ and rescaling $u,v$ by $\sqrt{\lambda_1},\sqrt{\lambda_2}$ we get
\be
  (\lambda_1 \lambda_2)^4 \int_{\C^4} \rd^4 u \, \rd^4 v \, |[u|v\ket|^4 \, \delta^{(4)}\left(  |u\ket\bra u| + |v\ket\bra v| - \lambda_1 \lambda_2|[u|v\ket|^2 \right)
\ee
Now defining 
\be
  \vec{U} = \bra u | \vec{\sigma} | u \ket \qquad \vec{V} = \bra v | \vec{\sigma} | v \ket
\ee
we can express $|u\ket\bra u| = (|\vec{U}|+\vec{U}\cdot \vec{\sigma})/2$ and $|u][ u| = (|\vec{U}|-\vec{U}\cdot \vec{\sigma})/2$ so
\be
  [v|u\ket\bra u | v] = \frac{|\vec{U}| |\vec{V}|-\vec{U} \cdot \vec{V}}{2}
\ee
Expressing the integrations over 3-vectors as in (\ref{eqn_spinor_spherical}) we have
\be
  (\lambda_1 \lambda_2)^4 \left(\frac{\pi}{4}\right)^2 \int \frac{\rd^3 \vec{U} \rd^3 \vec{V}}{|\vec{U}| |\vec{V}|} \left(\frac{|\vec{U}| |\vec{V}|-\vec{U} \cdot \vec{V}}{2}\right)^2 \delta^{(1)}\left(  |\vec{U}| + |\vec{V}| - (\lambda_1 \lambda_2)\frac{|\vec{U}| |\vec{V}|-\vec{U} \cdot \vec{V}}{2} \right) \delta^{(3)}\left( \vec{U} + \vec{V} \right)
\ee
Performing the integration over $\vec{V}$ 
\be
  (\lambda_1 \lambda_2)^4 \left(\frac{\pi}{4}\right)^2 \int \rd^3 \vec{U} \, |\vec{U}|^2 \, \delta^{(1)}\left(  2|\vec{U}| - \lambda_1 \lambda_2|\vec{U}|^2  \right)
\ee
Factoring out $\lambda_1\lambda_2 |\vec{U}|$ 
\be
  (\lambda_1 \lambda_2)^3 \left(\frac{\pi}{4}\right)^2 \int \rd^3 \vec{U} \, |\vec{U}| \, \delta^{(1)}\left(  \frac{2}{\lambda_1 \lambda_2} - |\vec{U}|  \right) 
\ee
and finally we identify $\text{Vol}(U(2)) = 4\pi^3$ 
\be
  (\lambda_1 \lambda_2)^3 \left(\frac{\pi}{4}\right)^2 \left( \frac{2}{\lambda_1 \lambda_2}\right)^3 4 \pi = \frac{1 }{2}\text{Vol}(\text{U}(2))
\ee

\section{Volume of the Grassmannian}
\label{app_vol_grass}

We wish to perform the integral
\be
\int_{\C^{2n}} \prod_{i=1}^{n} \rd^4 x_i \, \delta^{(4)}\left( \sum_{i=1}^{n} |x_i \ket \bra x_i | - 1 \right)
\ee
Define two $n$-dimensional complex vectors $\vec{A}$ and $\vec{B}$ by the two components of the $n$ spinors this integral becomes
\be
\int_{\C^{n}\times \C^{n}} \rd \vec{A} \, \rd \vec{B} \, \delta^{(4)} \bpm |\vec{A}|^2 -1 & \vec{A} \cdot \vec{B}^\ast \\ \vec{A}^\ast \cdot \vec{B} &  |\vec{B}|^2 -1 \epm 
\ee
Decomposing the delta in terms of the orthonormal basis 
$$
  \bpm \sqrt{2} & 0 \\ 0 & 0 \epm, \quad \bpm 0 & 1 \\ 1 & 0 \epm, \quad \bpm 0 & -i \\ i & 0 \epm, \quad \bpm 0 & 0 \\ 0 & \sqrt{2} \epm, \quad
$$
this integral becomes
\be
2 \int_{\C^{n}\times \C^{n}} \rd \vec{A} \, \rd \vec{B} \, \delta\left( |\vec{A}|^2-1 \right)\delta\left( |\vec{B}|^2-1 \right)\delta^{(2)}\left( \vec{A} \cdot \vec{B} \right)
\ee
Therefore this is two real $2n-1$ dimensional spheres with one complex condition $\vec{A} \cdot \vec{B} = 0$.  Performing a unitary transformation on $\vec{B}$ the complex delta becomes $\delta^{(2)}\left( B_1 \right)$. Thus this is the volume of a $2n-1$ dimensional sphere times the volume of a $2n-3$ dimensional sphere with an extra factor 1/4 from the Jacobians of $|\vec{A}|^2, |\vec{B}|^2$.  The volume of an $n$ dimensional sphere is $\text{vol}(S^{2n+1}) = (2\pi)\pi^{n}/n!$ thus
\begin{align}
\int_{\C^{2n}} \prod_{i=1}^{n} \rd^4 x_i \, \delta^{(4)}\left( \sum_{i=1}^{n} |x_i \ket \bra x_i | - 1 \right) &= \frac{1}{2}\text{vol}\left(S^{2n-1}\right) \text{vol}\left(S^{2n-3}\right) = \frac{1}{2} \text{vol}\left(\text{U}(2)\right) \text{vol}\left(\text{Gr}(2,n)\right) \nn \\ &= \frac{(4\pi^3)}{2} \frac{\pi^{2n-4}}{(n-1)!(n-2)!} 
\end{align}
where $\text{vol}\left(\text{Gr}(2,n)\right) = \pi^{2n-4}/(n-1)!(n-2)!$ which agrees with the formula given in \cite{Freidel:2010tt}.


\end{document}